# The electronic-structure origin of the anisotropic thermopower of nanolaminated $Ti_3SiC_2$ determined by polarized x-ray spectroscopy and Seebeck measurements

Martin Magnuson[1], Maurizio Mattesini[2,3], Ngo Van Nong[4], Per Eklund[1] and Lars Hultman[1]

[1]*Thin Film Physics Division, Department of Physics, Chemistry and Biology, IFM, Linköping University, SE-58183 Linköping, Sweden.*

[2]*Departamento de Física de la Tierra, Astronomía y Astrofísica I, Universidad Complutense de Madrid, Madrid, E-28040, Spain.*

[3]*Instituto de Geociencias (CSIC-UCM), Facultad de CC. Físicas, E-28040 Madrid, Spain.*

[4]*Department of Energy Conversion and Storage, Technical University of Denmark, Risö Campus, 4000 Roskilde, Denmark.*

## Abstract

Nanolaminated materials exhibit characteristic magnetic, mechanical, and thermoelectric properties, with large contemporary scientific and technological interest. Here, we report on the anisotropic Seebeck coefficient in nanolaminated $Ti_3SiC_2$ single-crystal thin films and trace the origin to anisotropies in element-specific electronic states. In bulk polycrystalline form, $Ti_3SiC_2$ has a virtually zero Seebeck coefficient over a wide temperature range. In contrast, we find that the in-plane (basal *ab*) Seebeck coefficient of $Ti_3SiC_2$, measured on single-crystal films has a substantial and positive value of 4-6 µV/K. Employing a combination of polarized angle-dependent x-ray spectroscopy and density functional theory we directly show electronic structure anisotropy in inherently nanolaminated $Ti_3SiC_2$ single-crystal thin films as a model system. The density of Ti 3*d* and C 2*p* states at the Fermi level in the basal *ab*-plane is about 40 % higher than along the *c*-axis. The Seebeck coefficient is related to electron and hole-like bands close to the Fermi level but in contrast to ground state density functional theory modeling, the electronic structure is also influenced by phonons that need to be taken into account. Positive contribution to the Seebeck coefficient of the element-specific electronic occupations in the basal plane is compensated by 73 % enhanced Si 3*d* electronic states across the laminate plane that give rise to a negative Seebeck coefficient in that direction. Strong phonon vibration modes with three to four times higher frequency along the *c*-axis than along the basal *ab*-plane also influence the electronic population and the measured spectra by the asymmetric average displacements of the Si atoms. These results constitute experimental evidence explaining why the average Seebeck coefficient of $Ti_3SiC_2$ in polycrystals is negligible over a wide temperature range. This allows the origin of anisotropy in physical properties of nanolaminated materials to be traced to anisotropies in element-specific electronic states.

# 1 Introduction

A *nanolaminate* is a material with a laminated - layered - structure in which the thicknesses of the individual layers are in the nanometer range. Inherently nanolaminated materials have a crystal structure containing alternating distinctly different layers within the unit cell [1]. As a consequence, inherently nanolaminated materials often exhibit particular mechanical [2], magnetic [3], or thermoelectric properties [4, 5]. Given their highly anisotropic structure, explaining the physical properties requires an in-depth understanding of the anisotropy and the orbital occupation [6] in the electronic structure. As a model system for inherently nanolaminated materials, we choose $Ti_3SiC_2$. This multifunctional metallic and ceramic compound [7, 8, 9] is interesting for this topic as it exhibits a number of fascinating properties that can be related to anisotropy. For example, the large anisotropy of the shear modulus of polycrystalline $Ti_3SiC_2$ reported in neutron diffraction is controversial as the experimental shear stiffness exceeds the theoretical predictions by a factor of three [10]. Further, the Seebeck coefficient (thermopower) of polycrystalline bulk $Ti_3SiC_2$ is negligible over a wide range of temperatures, a unique phenomenon that has broad implications on the understanding of the thermoelectric effect [11]. However, concrete evidence connecting the property anisotropies to the underlying electronic structure physics is missing [12, 13, 14, 15]. Ground state density functional theory studies shows that the band structure of $Ti_3SiC_2$ is quite unique as it has an electron-like and a hole-like band nestled around the Fermi level that have important effects for the transport properties [16, 17]. These theoretical results also suggest that the negligible Seebeck coefficient in polycrystalline $Ti_3SiC_2$ is due to an average cancellation between the partial Seebeck coefficient with positive sign along the main high symmetry axis in the basal plane with the negative Seebeck coefficient along the c-axis. However, the Seebeck coefficient is related to the conductivity effective mass that varies continuously with the band structure in all directions. Moreover, the model assumes isotropic scattering and carrier mobility (the Mott approximation), which is invalid for nanolaminates and contradicted by experimental determination of the anisotropy in carrier lifetimes [18].

In this paper, we show direct experimental evidence of anisotropic Seebeck coefficient and its relation to high anisotropy in the electronic structure in epitaxially grown single crystal thin films of $Ti_3SiC_2(0001)$. The in-plane (basal *ab*) Seebeck coefficient of $Ti_3SiC_2$, measured on single-crystal films has a substantial and positive value of 4-6 µV/K. This is in stark contrast to the virtually zero value for bulk polycrystals, which thus can be interpreted as due to compensation between the electron-like and hole-like bands, wrapped around the Fermi level, in and across the laminate plane. We further show the strong influence of phonons in the Si layers, supported by thorough *ab initio* band structure calculations. Surprisingly, an enhancement of the nominally hole-like states perpendicular to the laminate basal plane by occupied Si 3*d* states is observed and is traced to a different phonon mode that compensates for the larger electronic contribution of the electron-like Ti 3*d*, C 2*p* and Si 3*p* states along the laminate basal plane. Thus, the near-zero thermopower is a direct effect of anisotropy in the electronic structure at the conduction and valence band edges strongly influenced by phonons. Although earlier density functional theoretical calculations have provided a qualitative explanation of the negligible thermopower in $Ti_3SiC_2$ by compensating rigid anisotropic bands [16, 17], accounting for phonons are necessary for a more correct description of the band occupations.

# 2 Experimental details

## 2.1 Ti$_3$SiC$_2$(0001) thin film synthesis and thermoelectric characterization

The 500 nm thick Ti$_3$SiC$_2$(0001) thin film investigated here was deposited by dc magnetron sputter epitaxy (base pressure $\sim 6\times10^{-8}$ Pa) from elemental targets of Ti and C, and Si in an argon discharge at a pressure of 4 mTorr on an α-Al$_2$O$_3$(000$l$) substrate. The films are single-phase epitaxial Ti$_3$SiC$_2$ essentially identical to those described in Ref [12, 19].

The in-plane Seebeck coefficient was measured by static DC method using an ULVAC RIKO ZEM-3 thermoelectric property measurement system in a low pressure He atmosphere. The size of the specimen is $\sim 4\times10$ mm$^2$. Silver (Ag) paste was used to ensure the contacts of the film with electrodes and two measured probes (R-type thermocouples), which were pressed by a spring on the surface of the film during the measurement. V-I plot measurement was made to judge if the lead is in intimate contact with a set sample.

## 2.2 X-ray emission and absorption measurements

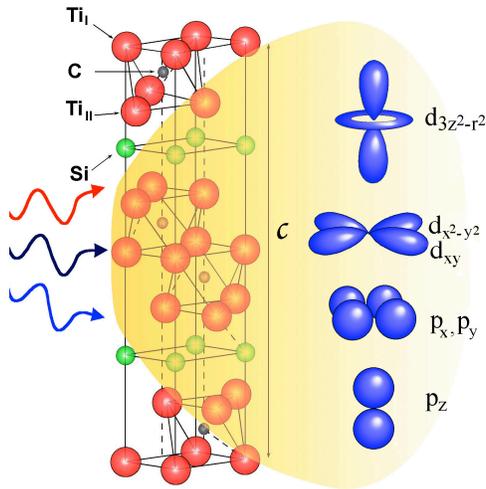

**Figure 1:** (Color online) Illustration of the hexagonal crystal structure of Ti$_3$SiC$_2$ and electronic orbitals of the chemical bonds across ($d_{3z^2-r^2}$, $p_z$) and in ($d_{xy}$, $d_{x^2-y^2}$, $p_x$, $p_y$) the laminate plane. The Ti atoms have two different sites, denoted Ti$_I$ and Ti$_{II}$ and every fourth layer of the TiC slabs is interleaved by a pure Si layer.

The orbital occupation and the electronic anisotropy was investigated using bulk-sensitive and element-specific soft x-ray fluorescence in absorption (SXA), emission (SXE), and resonant inelastic x-ray scattering (RIXS) to probe the unoccupied and occupied bands of the different elements. By tuning the energy of linearly polarized x-rays to the specific core levels of Ti, C, and Si and changing the incidence angle of the x-rays from grazing to near normal relative to the laminate plane, information about the occupied and unoccupied electronic orbitals across ($d_{3z^2-r^2}$, $p_z$) and in ($d_{x^2-y^2}$, $d_{xy}$, $p_x$, $p_y$) the laminate plane was obtained as illustrated in Figure 1.

The SXA and SXE spectra were measured at 15$^o$ and 75$^o$ incidence angles at 300 K and $\sim 1\times10^{-7}$ Pa at the undulator beamline I511-3 on the MAX II ring of the MAX IV Laboratory, Lund University, Sweden). The energy resolutions at the Ti 2*p*, C 1*s*, Si 2*s*, and Si 2*p* edges of the beamline monochromator were 1.6, 1.0, 0.2 eV and 0.01 eV and the SXE spectra were recorded with spectrometer resolutions 0.7, 0.2, 0.2 eV, and 0.01, respectively. The SXA spectra were measured in total fluorescence yield (TFY) both at $15^o$ (along the *c*-axis, near perpendicular to the basal *ab* plane) and $90^o$ (normal, parallel to the basal *ab*-plane) incidence angles and normalized by the step edge below and far above the absorption thresholds. The SXE spectra were measured at $15^o$ (in the basal *ab* plane) and $75^o$ (perpendicular to the basal *ab*-plane) incidence angles. For comparison of the spectral profiles, the measured SXE data were normalized to unity and were plotted on a common photon energy scale (top) and relative to the $E_F$ (bottom) using the $2p_{1/2}$ core-level XPS binding energy of 460.6 eV ($2p_{3/2}$=454.6 eV) for $Ti_3SiC_2$ measured on the same sample. Calculated nonresonant SXE spectra for the two Ti sites ($Ti_I$ and $Ti_{II}$) are also shown at the bottom of Figure 2, calculated C *K* SXE in Figure 3 and Si $L_1$, $L_{2,3}$ SXE in Figure 4.

# 3 Computational details

## 3.1 First-principles calculations

The *ab initio* calculations were performed according to the Density Functional Theory (DFT) [20] employing the Full Potential Linearized Augmented Plane Wave (FPLAPW) calculational scheme of the Wien2k code [21]. Electronic exchange-correlation effects were treated via the Generalized Gradient Approximation (GGA) as parameterized by Perdew, Burke, and Ernzerhof (PBE) [22]. A plane wave cut-off corresponding to $R_{MT}*K_{max}$=8 was used, while the charge density and potentials were expanded up to $\ell$=12 inside the atomic spheres. Total energy convergency was achieved with respect to the Brillouin zone (BZ) integration using a 15×15×3 mesh-division of the reciprocal lattice vectors, corresponding to 54 irreducible k-points.

For the computations of the SXE spectra, we used the so-called *final-state rule* [23], where no core-hole was created at the photo-excited atom. For SXA, however, we specifically included core-hole effects by means of the super-cell technique.

## 3.2 Calculation of x-ray absorption and emission spectra

The soft x-ray absorption spectra were computed within the same theoretical *ab initio* scheme used for emission spectra but with the inclusion of core-hole effects. Thus, we specifically considered a crystal potential created from a static screening of the core-hole by the valence electrons. We therefore generated such a self-consistent-field potential using a 2×2×1 hexagonal supercell of 48 atoms containing one core-hole on the investigated element. The electron neutrality of the system was maintained constant through a negative background charge. With this procedure we guarantee the treatment of the excitonic coupling between the screened core-hole and the conduction electrons, which is an important requisite for simulating SXA edges in semiconducting materials.

Theoretical emission spectra were computed within the single-particle transition model and using the electric-dipole approximation, which means that only the transitions between the core states with orbital angular momentum $\ell$ to the $\ell\pm 1$ components of the valence bands were considered. Core-hole lifetime and experimental broadening was used in accordance with the experiment. In the fitting procedure to the 490.0 eV Ti $L_{2,3}$ SXE spectra, we employed the experimental values for the $L_3/L_2$ ratio of 6:1 in the basal plane and 15:1 along the *c*-axis. We used the experimental $L_{2,3}$ peak splitting of 6.2 eV, which is slightly larger than our calculated *ab initio* spin-orbit splitting of 5.7 eV.

### 3.3 Phonon calculations

Phonon densities of states (PhDOS) were calculated by employing the so-called supercell approach in the framework of density functional theory and Density-Functional Perturbation Theory (DFPT) [24]. Specifically, a 2×2×1 supercell system was used to compute real space force constants within the q-ESPRESSO software package [25]. Phonon frequencies were then obtained from the force constants by using the QHA code [26]. The PBE generalized gradient approximation was chosen to treat the exchange-correlation effects. The ultrasoft pseudo-potentials in the form introduced by Vanderbilt [27] were used, while the integration over the BZ was performed by using the special k-points technique [28] with a Gaussian broadening of 0.002 Ry. The electron wave functions were expanded in the plane wave basis set with a cut-off of 70 Ry and plane wave kinetic energy up to 700 Ry.

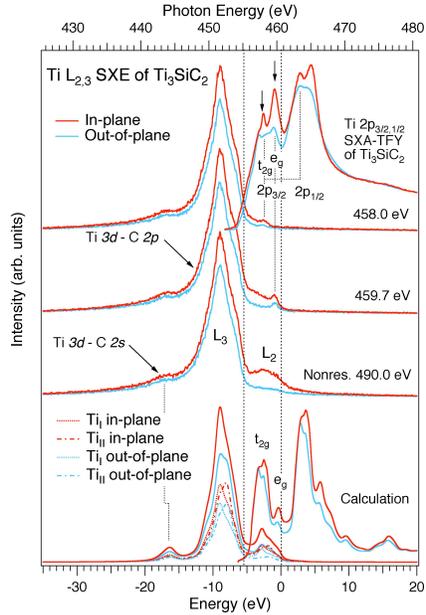

**Figure 2:** (Color online) Top-right: experimental Ti 2*p* SXA-TFY spectra of $Ti_3SiC_2$ following the $2p_{3/2,1/2} \rightarrow 3d$ dipole transitions. Left top-to-bottom: resonant Ti $L_{2,3}$ SXE spectra with excitation energies 458.0, 459.7 eV, indicated by the vertical arrows in the SXA spectra, and nonresonant spectra at 490.0 eV. Bottom, calculated spectra of the two different crystallographic Ti sites, $Ti_I$ and $Ti_{II}$ using the experimental spin-orbit peak-splittings and $L_3/L_2$ branching ratios.

## 4 Results

### 4.1 Ti $L_{2,3}$ x-ray absorption and emission

Figure 2 (top-right) shows Ti 2*p* SXA spectra of the 3*d* and 4*s* conduction bands in the basal *ab*-plane and along the *c*-axis with the main peak structures associated with the $2p_{3/2}$ and $2p_{1/2}$ core shells. It should be noted that the splitting energy between the two different crystallographic Ti sites denoted by $Ti_I$ and $Ti_{II}$ (Fig. 1) is negligible. The observed sub-peaks are predominantly related to the $t_{2g}$-$e_g$ splitting and the intensities of the peaks were found to be about 20% higher for the in-plane orbitals than for the out-of-plane ones. The minority $t_{2g}(d_{xy}, d_{xz}, d_{yz})$ orbitals have the lowest energy and the $e_g(d_{3z^2-r^2}, d_{x^2-y^2})$ orbitals are located at 1.7 eV higher energy in the non-purely octahedral crystal field around the Ti sites [29], consistent with our DFT calculations shown at the bottom-right of Fig. 2. The branching ratios that largely deviates from the statistical 2:1 ratio are due to the exchange and mixed terms between the core-states [30]. The intensity of the unoccupied states at the $E_F$ is 44 % higher in the basal ab-plane than across the laminate plane.

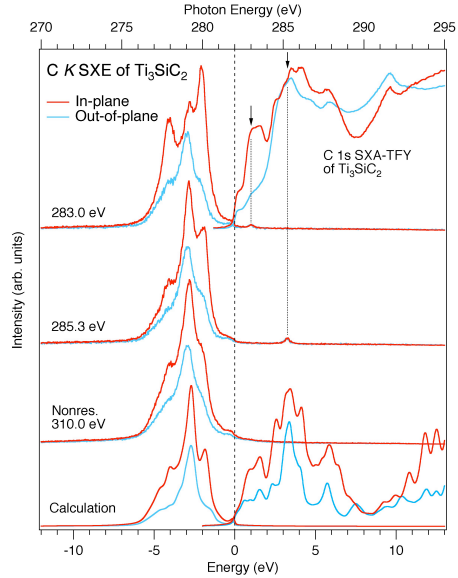

**Figure 3:** (Color online) Top-right: C 1s SXA-TFY spectra of $Ti_3SiC_2$. Left top-to-bottom: resonant spectra excited at 283.0 and 285.3 eV indicated by the vertical arrows and nonresonant C K SXE spectra at 310.0 eV. All spectra are plotted on a photon energy scale (top) and a relative energy scale (bottom) with respect to the top of the valence band using the C 1s core-level X-ray Photoelectron Spectroscopy (XPS) binding energy of 282.0 eV measured on the same sample.

Ti $L_{2,3}$ RIXS spectra are shown in the center of Fig. 2, probing the occupied Ti 3d and 4s valence bands of $Ti_3SiC_2$ in the basal ab-plane and along the c-axis at the $t_{2g}$ and $e_g$ absorption maxima [29] of the $2p_{3/2}$ core shell (vertical arrows in the SXA spectra) and nonresonant. The spectral shape of the main $L_3$ peak at -3.2 eV below $E_F$ reflects the Ti 3d - C 2p hybridization and orbital overlap and is similar for all excitation energies although the in-plane spectra have higher intensities. A weak shoulder at -10 to -12 eV below $E_F$ is due to Ti 3d - C 2s hybridization and is completely absent in pure Ti metal [29]. At the $E_F$, the intensity is 47 % higher in the laminate basal ab-plane than along the c-axis consistent with our DFT calculations. The calculated spectra indicate that the spectral weight at the $E_F$ is slightly higher for the $Ti_I$ atomic sites that are bonded to C only than for the $Ti_{II}$ atomic sites that are bonded to both C and Si atoms.

## 4.2 C K x-ray absorption and emission

Figure 3 (top right) shows experimental C 1s SXA spectra measured across and in the laminate ab-plane probing the C 2p conduction band. The measured intensity is two times higher for the hole-like states in the basal ab-plane than along the c-axis in the region 0-2 eV above $E_F$. This anisotropy is unexpectedly large, considering that the C atoms occupy the octahedral cavities defined by the three Ti layers in the $Ti_3C_2$ slabs between the Si monolayers (Fig. 1). The large anisotropy is consistent with our DFT calculations at the bottom-right in Fig. 3.

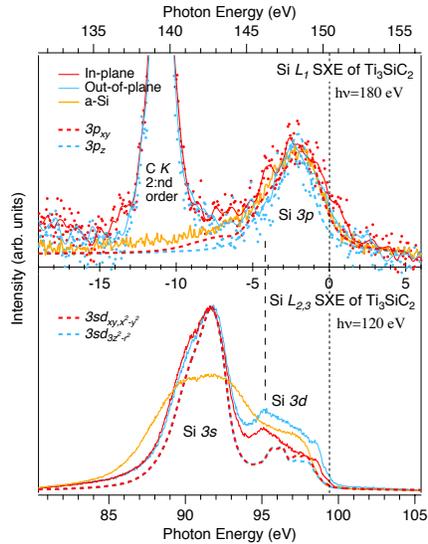

**Figure 4:** (Color online) Top panel: Si $L_1$ SXE spectra of $Ti_3SiC_2$ excited at 180 eV in comparison to pure amorphous Si (a-Si). The dashed curves are corresponding calculated $3p_{xy}$ and $3p_z$ spectra with an $L_{2,3}$ spin-orbit splitting of 0.646 eV and an $L_3/L_2$ ratio of 2:1. Bottom panel: Si $L_{2,3}$ SXE spectra excited at 120 eV in comparison to amorphous Si. The dashed curves are corresponding calculated $3sd_{xy,x^2-y^2}$ and $3sd_{3z^2-r^2}$ spectra. A common energy scale with respect to the top of the valence band edge (vertical dotted line) is indicated between the top and bottom panels.

RIXS and nonresonant C $K$ SXE spectra measured across and in the laminate $ab$-plane are shown on the left part in figure 3, corresponding to the occupied C $2p$ valence band of $Ti_3SiC_2$, in comparison to calculated nonresonant spectra. The spectra measured in the basal $ab$-plane exhibit a remarkable excitation-energy dependence while the spectral probed along the $c$-axis have a similar shape independent of excitation energy. The first in-plane spectrum excited at 283.0 eV exhibits three $2p_{xy}$ peaks with σ character at -3.8, -2.5, and -1.8 eV below $E_F$ corresponding to three bands in different crystallographic directions in the $ab$-plane. On the contrary, the spectra probed along the $c$-axis are dominated by a single $2p_z$ band with π character at -2.5 eV below $E_F$ with a shoulder on each side. The intensity of the shoulders decreases with increasing excitation energy both for spectra measured in the basal $ab$-plane and along the $c$-axis. The peak structures in the basal $ab$-plane and along the $c$-axis results from C $2p$ - Ti $3d$ hybridization and orbital overlap in the valence bands that are also influenced by the Si $3p$, $3d$ and $3s$ bands. For the nonresonant C $K$ spectra, the intensity of the occupied C $2p$ bands at the $E_F$ is 37 % higher in the basal $ab$-plane than along the $c$-axis consistent with the calculated spectra.

### 4.3 Si $L_1$ and $L_{2,3}$ x-ray emission

Figure 4 (top panel) shows experimental Si $L_1$ SXE spectra of $Ti_3SiC_2$ measured nonresonantly at 180 eV photon energy, probing the relatively weak Si $3p$ valence band across and in the laminate basal $ab$-plane (Fig. 1). For comparison, calculated spectra (dashed lines) and an isotropic spectrum of amorphous Si are also included.

The shape of the Si 3*p* valence band between 0 and -6 eV is a result of C 2*p* - Ti 3*d* hybridization and orbital overlap. The in-plane $3p_{xy}$ states containing three σ orbitals are spread out between 0 and -5 eV below $E_F$ while the $3p_z$ states with a single bonding π orbital are more localized around -2 eV below $E_F$. Experimentally, the Si 3*p* intensity is about 10 % higher in the basal *ab*-plane than along the *c*-axis at the $E_F$, while in the calculated spectra the anisotropy of the Si 3*p* states is much larger at the $E_F$ (70 %).

The bottom panel in Fig. 4 shows experimental Si $L_{2,3}$ SXE spectra of Ti$_3$SiC$_2$ aligned to the $2p_{1/2}$ core level binding energy at 99.4 eV ($2p_{3/2}$=98.9 eV) in comparison to isotropic amorphous Si excited nonresonantly at 120 eV. Comparing the Si $L_{2,3}$ SXE spectrum of Ti$_3$SiC$_2$ to the broad spectrum of amorphous Si, the main peak at -7.5 eV at the bottom of the valence band is attributed to an isotropic Si 3*s* band. The peak of the 3*s* band in Ti$_3$SiC$_2$ is produced at a symmetry point in the band structure by the orbital overlap with the Ti 3*d*-band, that is supported by our calculated Si $L_{2,3}$ SXE spectra (dashed lines). On the contrary, the upper part of the valence band in Ti$_3$SiC$_2$ between 0 and -6 eV is dominated by the Si 3*d* character weighting in the partial *DOS* that also participate in the Ti$_{II}$-Si bonding in Ti$_3$SiC$_2$.

The most interesting experimental observation is that the intensity of the Si 3*d* character weighting in the partial *DOS* in the upper valence band (0 to -6 eV) is much higher (73 % at the $E_F$) along the *c*-axis than in the basal *ab*-plane. This effect is significantly larger and completely opposite from that in Si $L_{2,3}$ spectra of pure single crystal Si(111) [9, 31]. The enhanced Si 3*d* (electron-like) intensity along the *c*-axis of Ti$_3$SiC$_2$ should reduce and compensate for the dominating hole-like Ti 3*d* bands that give negative contributions to the Seebeck coefficient along the *c*-axis. In particular, it should be noted that this strong intensity of the Si 3*d* character weighting in the partial *DOS* observed experimentally along the *c*-axis at room temperature is not at all reproduced in ground state DFT theory at zero Kelvin but is instead opposite in the energy range 1-3 eV below $E_F$. Variations in the electronic *DOS* in the vicinity of the $E_F$ is an indication of development of significant electron-phonon interaction that is the case for the *A*-atoms in $M_{n+1}AX_n$-phases as observed in neutron diffraction experiments [33]. Therefore, phonons must be included in the theory in order to at least qualitatively reproduce the Si spectra.

### 4.4 Phonon vibrations

Fig. 5 (top panel) shows calculated phonon frequency spectra in the *ab*-basal plane (Si-*x*, Si-*y*) and along the *c*-axis (Si-*z*). As observed, there are two different overlapping regimes of phonon vibration modes with an out-of-plane high-frequency peak three to four times higher in frequency (10-12 THz) than for the peak maximum for the in-plane vibrational mode (3.3 THz), consistent with previous calculations [34]. From neutron diffraction studies of $M_{n+1}AX_n$-phases, the *A*-atoms (Si in Ti$_3$SiC$_2$) are known to act as 'rattlers' following a rather complicated ellipsoidal phonon trajectory [33].

The bottom panel of Fig. 5 shows the Si 3*d* SXE character weighting in the partial *DOS* in Ti$_3$SiC$_2$ when the core-exited Si atoms are displaced along the *xy* basal plane (in-plane displacement) and along the *c*-axis (out-of-plane displacement). We applied the *static displacement method* to obtain the Si $L_{2,3}$ SXE spectra along both the $xy$- and the $z$-Si directions with a 2×2×1 supercell. Within a qualitative approach, the applied atomic displacements are approximations equivalent of electron-phonon calculations. We find that static displacement along the $c$-axis give rise to substantial anisotropy ($xy+x^2+y^2$ vs. $3z^2-r^2$) in the Si $L_{2,3}$ SXE spectra within 1 eV from the $E_F$, whereas displacement in the basal $ab$-plane results in much smaller anisotropy further away from the $E_F$. In particular, a sharp double peak structure near the $E_F$, with $d_{3z^2-r^2}$ character remains equally intense for both positive (+$z$) and negative (-$z$) displacement of the core-excited Si atom along the $c$-axis. This peak structure is absent not only for all orbitals with in-plane displacement of the core-excited Si atoms but also for the $xy+x^2+y^2$ orbitals with out-of-plane displacement. It shows that phonons must be included for the Si atoms to reproduce the spectra while the other atoms have lower phonon coupling. The shape of the Si $L_1$ spectra (Fig. 4, top) are in a similar way modified by displacement of the Si atoms. Displacement along the $c$-axis (+$z$) or (-$z$) reduces the intensity of the $3p_z$ band at -2 eV and increases the intensity in the vicinity of the $E_F$ while $xy$-displacement affects the spectra very little. For other systems with rapidly changing *DOS* at the $E_F$, it is known that significant electron-phonon interactions can give rise to anomalous temperature-dependent properties such as closing or opening of band gaps [35].

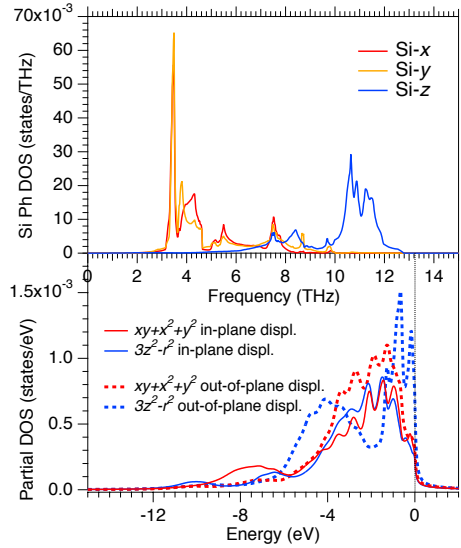

**Figure 5:** (Color online) Top panel: Calculated phonon frequency spectra (PhDOS) of the Si atoms in $Ti_3SiC_2$. Bottom panel: Calculated Si $3d$ spectra of $Ti_3SiC_2$ for in-plane ($xy$-basal plane) displacement and displacement along the $c$-axis (out-of-plane). The selected core-excited Si atom has been moved either along the $xy$-plane or perpendicular to it. The applied in-plane displacements correspond to 0.615 Å along both $x$ and $y$ crystallographic axis, giving a total displacement vector modulus of 0.870 Å. The out-of-plane displacement vector modulus is 0.884 Å along the $z$-axis.

## 4.5 Seebeck coefficients

Fig 6 shows measured and calculated Seebeck coefficients for $Ti_3SiC_2$. The in-plane measurements (i.e., of $S_{xx}$) on the present single-crystal thin films show substantial and positive values increasing essentially linearly from 4.4 µV/K at room temperature to 6.3 µV/K at 700 $^o$C. This is in stark contrast to the essentially zero value of bulk polycrystalline $Ti_3SiC_2$ (bottom curve in Fig. 6, from ref. [11]), thus providing direct evidence of the anisotropy of the Seebeck coefficient of $Ti_3SiC_2$. It is worth pointing out here that a direct measurement along the $c$-axis (i.e., of $S_{zz}$) is not possible on any existing single-crystal samples; thin films of this orientation do not exist (c.f., the discussion in Refs. [12] and [36]) and existing bulk single crystals [37, 38] are flat and too thin (~10 µm) to allow the necessary temperature gradient for a Seebeck measurement to be applied.

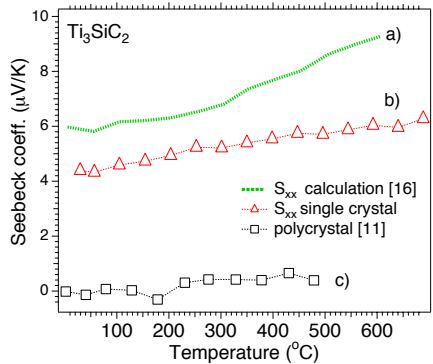

**Figure 6:** (Color online) Calculated and measured Seebeck coefficients of $Ti_3SiC_2$. a) $S_{xx}$ calculation [16], b) $S_{xx}$ measurement (present work), c) polycrystal $Ti_3SiC_2$ [11].

The calculated Seebeck coefficient $S_{xx}$ (top curve in Fig. 6, from ref [16]) overestimates the experimental value by ~25% at room temperature and the overestimate increases with temperature, reaching well over 50% at 600 $^o$C. This shows that a rigid band-structure calculation alone cannot fully explain the anisotropy and compensation in the Seebeck coefficient of $Ti_3SiC_2$. In order to provide a better explanation, the effect of phonons must be accounted for.

## 5 Discussion

Our results give direct proof of anisotropic Seebeck coefficient in single-crystal $Ti_3SiC_2$. Quantitatively, the measured values are 25% lower than theoretically estimated at room temperature and the difference increases strongly with temperature. The results show that the contribution from phonon vibrations from the Si atoms is significant in the anisotropic electronic structure and should thus affect the partial Seebeck coefficient with negative sign along the $c$-axis. The nature of the enhanced out-of-plane Si $L_{2,3}$ intensity observed experimentally can also be traced to the anisotropic band structure with an electron-like band (band 56) that encloses and shares the Fermi level together with the hole-like band (band 55) [16]. The electron-like band that dominates along the $c$-axis gives a negative sign to the Seebeck coefficient in that direction. Band structure calculations indicate that the 3$d$ states of the Ti atoms dominate at the Fermi level while the 2$p$ states of the C atoms and the 3$s$ and 3$p$ states of the Si atoms play a minor role for the transport properties. The behavior is largely related to the band filling in $Ti_3SiC_2$ and it is unique with negative Seebeck coefficient along the $c$-axis for a large temperature range. This has been correlated with the contribution of specific sheets of the Fermi surface at the corners of the Brillouin zone that depends on the exact $p$-band filling. This kind of anisotropic band structure compensation could likely also exist for other nanolaminates that has a negative contribution to the Seebeck coefficient along the $c$-axis, but the compensation by the band structure in $Ti_3SiC_2$ is unique. For comparison, the isostructural material $Ti_3GeC_2$ also has electron and hole-like bands wrapped around the Fermi level but with less symmetric band structure compared to $Ti_3SiC_2$ and has a negative contribution to the Seebeck coefficient along the *c*-axis, but the compensation is not complete. On the contrary, using Al atoms, $Ti_3AlC_2$ has lower *p*-band filling that shifts the nestled bands far above the Fermi level. This implies absence of specific sheets at the Fermi surface and positive Seebeck coefficient in all directions [41]. However, as shown here, not only the anisotropic band structure and band filling play a role for the transport properties, but also the coupled intrinsic motion of the constituent atoms.

Neutron diffraction experiments of $Ti_3SiC_2$ showed that the Si atoms act as rattlers and vibrate with an anisotropic elliptical thermal motion with the highest amplitudes in the basal plane simultaneously coupled with the $Ti_{II}$ atoms [33]. From the static displacement calculations of the core-excited Si atoms, we find that both positive and negative displacement along the *c*-axis significantly affects the anisotropy with enhanced intensity of the $d_{3z^2-r^2}$ orbitals close to the $E_F$. As the *z*-component of the Si phonon frequency also has 3-4 times higher frequency than the in-plane frequency, it implies that the electron-phonon coupling will mostly affect the top part of the valence band. The Si 2*p* core-hole decay in the SXE process has a lifetime of about 30 fs (20 meV) [39] that is on a faster time scale compared to the phonon vibrations i.e., about 80 fs out-of-plane and 300 fs along the basal plane. Our atomic displacement calculations of Si atoms rattling with high frequency along the *c*-axis result in an enhancement of the Si 3*d* SXE intensity within 0-5 eV as observed in the experiment. On the contrary, displacement along the laminate *ab*-plane gives less anisotropy in the Si 3*d* states observed experimentally due to the lower frequency. As the electronic emission process of ~ 30 fs is 10 times faster than the in-plane (Si-*x*, Si-*y*) phonon frequency, mostly the out-of-plane (Si-*z*) phonon branches affect the spectral line shape. Therefore, the in-plane atomic motions can be seen as more or less ÔfrozenÕ during the SXE process, thus keeping mostly the out-of-plane effects in the spectral distribution. This implies that, when averaged out, the largest contribution to the spectral change originates from the *c*-axis rattling that mainly affects the 0-5 eV region of the total Si $L_{2,3}$ SXE spectrum. Therefore, the electron-phonon coupling in $Ti_3SiC_2$ have an observable effect on the electronic structure of the Si atoms by modifying the screening properties of the electron density, particularly for the Si 3*d* character weighting in the partial *DOS* states. This effect should also influence the temperature-dependence of the Seebeck coefficient.

Thus, not only the anisotropic band structure and the constituent elements, but also the intrinsic motion of the Si atoms determines the anisotropy of the electronic structure. Band structure calculations at zero temperature do not sufficiently describe the materials properties as not only the Ti 3*d* states play an important role in the band structure but also the Si 3*d* character weighting in the partial *DOS* when phonons are taken into account. Our results for *single-crystal* $Ti_3SiC_2$ show experimentally that there is substantial anisotropy in the electronic structure that not only depends on the occupation of the bands but also by substantial phonon vibrations in different crystal orientations that must be included in the theory. This gives positive and negative contributions to the Seebeck coefficient in different directions given by the sign of the respective charge carriers. Thus our results fundamentally advance the understanding of the thermoelectric effect in anisotropic crystals, and allow these effects to be traced to anisotropies in element-specific electronic states. This approach can generally be applied on any system to correlate the properties to the electronic structure.

# 6 Conclusions

Our investigation of single-crystal $Ti_3SiC_2$(0001) demonstrates the definite anisotropy in the material's electronic structure thus resolving a long-standing research issue. The density of states in the laminate basal *ab*-plane is much larger than that along the *c*-axis for the Ti 3*d* and the C 2*p* states. The contribution to the anisotropic Seebeck coefficient of these occupations in the basal *ab*-plane are compensated by a relative enhancement of the Si 3*d* character weighting in the partial *DOS* perpendicular to the laminate basal *ab*-plane, and *vise versa*. This is due to the anisotropic electron-phonon interactions in the two crystallographic directions with different frequencies. Thus, the near-zero thermopower in polycrystals is a direct effect of anisotropy in the electronic structure. Rattling with higher frequency of Si atoms along the *c*-axis implies substantial anisotropy between the in-plane and out-of-plane orbitals that need to be included in theoretical modeling, as shown in this work.

# 7 Acknowledgements

We thank the Swedish Research Council (VR) LiLi-NFM Linnaeus Environment and project Grant No. 621-2009-5258, the staff at the MAX IV laboratory for experimental support and K. Buchholt, Linköping University, for providing the $Ti_3SiC_2$ sample.

# References

[1]    D. Music and J. M. Schneider; JOM **59**, 60 (2007).
[2]    Y. Gogotsi, A. Nikitin, Y. H. Ye, W. Zhou, J. E. Fischer, B. Yi, H. C. Foley and M. W. Barsoum; Nature Materials **2**, 591 (2003).
[3]    J. Mannhart and D. G. Schlom; Science **327**, 1607 (2010).
[4]    J. P. Heremans, V. Jovovic, E. S. Toberer, A. Saramat, K. Kurosaki, A. Charoenphakee, S. Yamanaka and G. J. Snyder; Science **321**, 554 (2008).
[5]    G. J. Snyder and E. S. Toberer; Nature Materials **7**, 105 (2008).
[6]    E. Benckiser, M. W. Haverkort, S. Bruck, E. Goering, S. Macke, A. Frano, X. Yang, O. K. Andersen, G. Cristiani, H.-U. Habermeier, A. V. Boris, I. Zegkinoglou, P. Wochner, H.-J. Kim, V. Hinkov and B. Keimer; Nature Materials **10**, 189 (2011).
[7]    M. W. Barsoum; Prog. Solid State Chem. **28**, 201 (2000).
[8]    J. Wang and Y. Zhou; Annu. Rev. Mater. Res. **39**, 415 (2009).
[9]    M. Magnuson, J. -P. Palmquist, M. Mattesini, S. Li, R. Ahuja, O. Eriksson, J. Emmerlich, O. Wilhelmsson, P. Eklund, H. Högberg, L. Hultman, U. Jansson; Phys. Rev. B **72**, 245101 (2005).
[10]   E. H. Kisi, J. F. Zhang, O. Kirstein, D. P. Riley, M. J. Styles and A. M. Paradowska; J. Phys. Condens. Mat. **22**, 162202 (2010).
[11]   H.-I. Yoo, M. W. Barsoum and T. El-Raghy; Nature **407**, 581 (2000).
[12]   P. Eklund, M. Beckers, U. Jansson, H. Hogberg, L. Hultman; Thin Solid Films **518**, 1851 (2010).
[13]   T. H. Scabarozi, P. Eklund, J. Emmerlich, H. Hogberg, T. Meehan, P. Finkel, M.W. Barsoum, J.D. Hettinger, L. Hultman, S.E. Lofland; Solid State Commun. **146**, 498 (2008).
[14]   N. Haddad, E. Garcia-Caurel, L. Hultman, M. W. Barsoum, and G. Hug; J. Appl. Phys. **104**, 023531 (2008).
[15]   V. Mauchamp, G. Hug, M. Bugnet, T. Cabioch and M. Jaouen; Phys. Rev. B **81**, 035109 (2010).
[16]   L. Chaput, G. Hug, P. Pecheur, and H. Scherrer; Phys. Rev. B **71**, 121104 (2005).
[17]   L. Chaput, P. Pecheur and H. Scherrer; Phys. Rev. B **75**, 045116 (2007).

[18] G. Hug, P. Eklund and A. Orchowski; Ultramicr. **110**, 1054 (2010).
[19] J. Emmerlich, J.-P. Palmquist, H. Hogberg, J. M. Molina-Aldareguia, Z. Czigany, S. Sasvari, P. Persson, U. Jansson and L. Hultman; J. Appl. Phys. **96**, 4817 (2004).
[20] W. Kohn and L. J. Sham; Phys. Rev. **140**, A1133 (1965).
[21] P. Blaha, Schwarz K, Madsen G K H, Kvasnicka D and Luitz J, (2001) *WIEN2K*,, An Augmented Plane Wave + Local Orbitals Program for Calculating Crystal Properties (Karlheinz Schwarz, Techn. Universit→t Wien, Austria). ISBN 3-9501031-1-2.
[22] J. P. Perdew, S. Burke and M. Ernzerhof; Phys. Rev. Lett., **77**, 3865 (1996).
[23] U. von Barth U and G. Grossmann; Phys. Rev. B **25** 5150 (1982) .
[24] S. Baroni, S. de Gironcoli, A. Dal Corso and P. Giannozzi; Rev. Mod. Phys. **73**, 515 (2001).
[25] Quantum-ESPRESSO, see http://www.quantum-espresso.org and http://www.pwscf.org.
[26] E. Isaev, QHA project, http://qe-forge.org/qha.
[27] D. Vanderbilt; Phys. Rev. B **41** 7892 (1990).
[28] H. J. Monkhorst and J. D. Pack; Phys. Rev. B **13**, 5188 (1976).
[29] M. Magnuson, E. Lewin, L. Hultman, and U. Jansson; Phys. Rev. B **80**, 235108 (2009).
[30] R. Laskowski and P. Blaha; Phys. Rev. B **82**, 205104 (2010).
[31] P. O. Nilsson, J. Kanski, J. V. Thordson, T. G. Andersson, J. Nordgren, J. Guo, and M. Magnuson; Phys. Rev. B **52**, R8643 (1995).
[32] J.-J. Gu, D. Zhang and Q. X. Guo; Solid State Commun. **148**, 10 (2008).
[33] N. Lane, S. C. Vogel and M. W. Barsoum; Phys. Rev. B **82**, 174109 (2010).
[34] A. Togo, L. Chaput, I. Tanaka and G. Hug; Phys. Rev. B **81**, 174301 (2010).
[35] O. Delairea, K. Martya, M. B. Stonea, P. R. C. Kenta, M. S. Lucasb, D. L. Abernathya, D. Mandrusa, and B. C. Salesa; Proc. Nat. Acad. Sci. **108**, 4725 (2011).
[36] P. Eklund, M. Bugnet, V. Mauchamp, S. Dubois, C. Tromas, J. Jensen, L. Piraux, L. Gence, M. Jaouen, T. Cabioch, Phys. Rev. B **84**, 075424 (2011).
[37] F. Mercier, O. Chaix-Pluchery, T. Ouisse, D. Chaussende, Appl. Phys. Lett. **98**, 081912 (2011).
[38] F. Mercier, T. Ouisse, D. Chaussende, Phys. Rev. B **83**, 075411 (2011).
[39] P. D. Padova, R. Larciprete, C. Quaresima, C. Ottaviani, B. Ressel and P. Perfitti; Phys. Rev. Lett. **81**, 2320 (1998).
[40] A. Sakai, T. Kanno, S. Yotsuhashi, A. Odagawa and H. Adachi; Japanese J. Appl. Phys. **44**, L966 (2005).
[41] L. Chaput, G. Hug, P. Pecheur, and H. Scherrer; Phys. Rev. B **75**, 035107 (2007).